\PassOptionsToPackage{table}{xcolor}
\documentclass[sigconf,authorversion=true]{acmart}

\usepackage{makecell}
\usepackage{tabularx}
\usepackage{longtable}
\usepackage[toc,page]{appendix}
\AtBeginDocument{%
  }

\copyrightyear{2022} 
\acmYear{2022} 
\setcopyright{rightsretained} 
\acmConference[CCS '22]{Proceedings of the 2022 ACM SIGSAC Conference on Computer and Communications Security}{November 7--11, 2022}{Los Angeles, CA, USA}
\acmBooktitle{Proceedings of the 2022 ACM SIGSAC Conference on Computer and Communications Security (CCS '22), November 7--11, 2022, Los Angeles, CA, USA}
\acmDOI{10.1145/3548606.3564253}
\acmISBN{978-1-4503-9450-5/22/11}

\begin{document}

\title[Patient Community: A Test Bed For Privacy Threat Analysis]{Poster: Patient Community -- \\ A Test Bed For Privacy Threat Analysis}

%\email{webmaster@marysville-ohio.com}
%\affiliation{%
%  \institution{Institute for Clarity in Documentation}
%  \streetaddress{P.O. Box 1212}
%  \country{USA}

\author{Immanuel Kunz}
\email{immanuel.kunz@aisec.fraunhofer.de}
\affiliation{%
  \institution{Fraunhofer AISEC}
  \city{Garching b. M\"unchen}
  \country{Germany}
}
\author{Angelika Schneider}
\email{angelika.schneider@aisec.fraunhofer.de}
\affiliation{%
  \institution{Fraunhofer AISEC}
  \city{Garching b. M\"unchen}
  \country{Germany}
}
\author{Christian Banse}
\email{christian.banse@aisec.fraunhofer.de}
\affiliation{%
  \institution{Fraunhofer AISEC}
  \city{Garching b. M\"unchen}
  \country{Germany}
}
\author{Konrad Weiss}
\email{konrad.weiss@aisec.fraunhofer.de}
\affiliation{%
  \institution{Fraunhofer AISEC}
  \city{Garching b. M\"unchen}
  \country{Germany}
}
\author{Andreas Binder}
\email{andreas.binder@aisec.fraunhofer.de}
\affiliation{%
  \institution{Fraunhofer AISEC}
  \city{Garching b. M\"unchen}
  \country{Germany}
}

\renewcommand{\shortauthors}{Kunz et al.}

% Call for posters:
% Posters are solicited that present unpublished in-progress or late-breaking research or extensions of published work on security-and privacy-related topics. Authors are asked to submit a short proposal that describes the main contributions of the poster or demonstration. Proposals should contain a brief abstract, place an emphasis on the motivation for the work, and summarize contributions being presented. Preliminary results may also be included. Presentation proposals will be evaluated primarily on their potential to stimulate interesting discussions, facilitate the exchange of ideas, and promote collaborations.
% TODO: highlight in-progress (still extending it, evaluating tools), highlight problem motivation of privacy in code and deployment info, hint at collaborations (tool developers, AI data generation, privacy engineers, ...) 

\begin{abstract}
Research and development of privacy analysis tools currently suffers from a lack of test beds for evaluation and comparison of such tools.
In this work, we propose a benchmark application that implements an extensive list of privacy weaknesses based on the LINDDUN methodology. It represents a social network for patients whose architecture has first been described in an example analysis conducted by one of the LINDDUN authors. We have implemented this architecture and extended it with more privacy threats to build a test bed that enables comprehensive and independent testing of analysis tools. % TODO , and can serve as a basis for discussion about privacy threats on the code- and deployment-level.
\end{abstract}




\maketitle

% TODO threat vs weakness vs vulnerability
\section{Introduction}
Various factors are increasing the pressure to automate privacy and security methods. These include iterative development methods, like agile, dynamic deployment environments, like the cloud, as well as the increasing size of applications.

For security analysis, there is a large number of methods and tools for the analysis of applications (see e.g.~\cite{zhang2021efficiency}) as well as respective benchmarks and data sets. For static application security testing (SAST), for instance, there exist specialized data sets such as the Juliet Test Suite~\cite{black2018juliet}. There is also the Damn Vulnerable Web Application project (DVWA)~\cite{dvwad} maintained by OWASP, which lists a number of projects that implement various vulnerabilities in different application domains.
Developing tools and methods for revealing privacy weaknesses in source code and deployment environments, however, remains a sparsely investigated area of research. One reason for this gap is that useful benchmarks are missing. Moreover, there is little discussion in general about how privacy threats materialize in actual source code and deployments.

In this paper, we present a benchmark application that implements and documents a collection of privacy threats: the Patient Community (PC) social network. It is based on a privacy threat model developed by Wuyts~\cite{Wuyts} which describes the architecture of the application and lists a number of privacy threats. We have implemented this application including most of the underlying weaknesses identified by Wuyts and the ones defined in the LINDDUN GO framework~\cite{wuyts2020linddun}.

With this work we aim to provide a test bed that encourages researchers and practitioners to develop and test privacy analysis tools, use it for educational purposes, as well as a basis for discussion about the code- and deployment-level analysis of privacy weaknesses.

\section{Background and Related Work}
\subsection{Privacy Threat Modeling}
% TODO STRIDE citation ~\cite{potter2009microsoft}
LINDDUN~\cite{deng2011privacy} is an acronym for privacy threats---linkability, identifiability, non-repudiation, detectability, disclosure, unawareness, and policy non-compliance---similar to STRIDE in security. Other works have partly proposed different privacy threats or respective protection goals~\cite{hansen2015protection,223961}. In this paper, we use the LINDDUN threats since they are more granular than other proposals. % TODO and cover them as well.

LINDDUN GO~\cite{wuyts2020linddun} covers the same threats as the original LINDDUN method, but is a more recent version that presents the threats in a consolidated form, which makes it easier to discuss and apply the method. Note that LINDDUN GO does not neglect any threats but simply presents them on a more abstract level: It is organized in the five LINDDUN categories (see above), which each are further divided into 5-7 more concrete threat types, e.g. Detectability includes \textit{Detectable credentials}, \textit{Detectability at storage}, and others. 
 % LINDDUN GO improves the usability and overview of the possible threats, while the authors claim that it does not omit any threats from the previous LINDDUN version. 

% The term \textit{threat} and its related concepts \textit{weakness} and \textit{vulnerability} are sometimes interpreted in different ways. 
In this paper, we understand a threat as a way to exploit an existing weakness in a system. For example, an API that leaks personal data is a weakness, while a threat would refer to an attacker finding and abusing this API to obtain the personal data. The focus of this paper is therefore the implementation of \textit{privacy weaknesses} which can be exploited by potential threats.

% FIXME: Why is this here? How is this relevant? Do you not want to say that what is missing is a common testbaed in order to compare the works?
% \subsection{Security and Privacy Analysis}
\subsection{Tools and Benchmarks}
Many tools exist to automatically analyze applications for potential security threats (see Zhang et al.~\cite{zhang2021efficiency}) which can be classified, for example, as black box~\cite{jung2008privacy} and white box~\cite{enck2014taintdroid,chen2015droidjust} approaches. In the area of data protection, there are also approaches for automated policy analysis~\cite{harkous2018polisis}. Also, semi-automated approaches exist which use extended data flow diagrams to facilitate security and privacy analysis. 
Furthermore, we have proposed a graph-based tool for semi-automated privacy threat analysis~\cite{kunz2022ppg}. Yet, there are few automated privacy analysis tools available. 

For all such approaches, realistic test beds and data sets are important to facilitate their evaluation and comparison. 
The \textit{Damn Vulnerable Web Applications Directory} by OWASP~\cite{dvwad} lists an extensive set of test applications which exhibit security vulnerabilities. Furthermore, testing suites for mobile security analysis have been proposed, e.g. by Arzt et al.~\cite{arzt2014flowdroid}.
For static privacy testing, we have recently proposed a testing library that implements 22 privacy weaknesses~\cite{kunz2022ppg}. These test cases are self-contained implementations that use mock configurations to mock databases and deployment environment. 
% A complete application, however, is not available
% The real-world applicability of such self-contained threats, however, is limited
In contrast, we present a complete application in this paper which is more realistic and more complex to analyze: it includes authentication, authorization and anonymization functionalities, as well as real databases and dynamic configurations. To the best of our knowledge, there are no comparable test beds available.

% \begin{itemize}
%     \item Many tools exist to analyze applications for potential security threats automatically: Static analysis, e.g. cpg, % TODO add references from reviews
%     % TODO review "whitebox" approaches again
%     \item There are various possibilities to classify security and privacy analysis tools, e.g. black box~\cite{jung2008privacy} and white box~\cite{enck2014taintdroid,chen2015droidjust} tools. In the area of data protection, there is are also approaches for automated policy analysis~\cite{harkous2018polisis}. % (which is based on the labeled policy corpus proposed by Wilson et al.~\cite{wilson2016creation}) % PPG % TODO PPG preprint to be published
%     % Also, static and dynamic analysis can be differentiated.
%     \item Extended data flow diagrams
%     \item threatspec
%     \item In summary, there are few automated privacy analysis tools, especially outside the mobile domain.
% \end{itemize}

% \subsection{Benchmarks and Test Beds}

% \item \cite{hao2019constructing}
% \begin{itemize}  
%     \item For static security testing, the academic literature often focuses on certain benchmarks, e.g. WU-FTPD, BIND, and others % TODO rather old reference Zitser et al. testing static analysis tools... ask Konrad or Alex?
%     % TODO remove recently for camera-ready
% \end{itemize}

In summary, security analysis and respective benchmarks are well researched and maintained, while there is little research into privacy-related tooling and benchmarks. To the best of the authors' knowledge, an application with real deployment configurations as a privacy benchmark has not been proposed before.

\section{Implementation}
% \subsection{Goals}
% \label{sec:goals}

In general, we follow two goals in our implementation: First, we aim at covering as many types of privacy threats as possible (in a static implementation) defined by LINDDUN GO~\cite{wuyts2020linddun} and second, we aim at including a diverse set of technologies, e.g. different programming languages, to prevent bias on a specific technology.

\subsection{Architecture and Components}
\label{sec:archandcomps}

\begin{figure}
    \begin{center}
        \includegraphics[width=\linewidth, keepaspectratio]{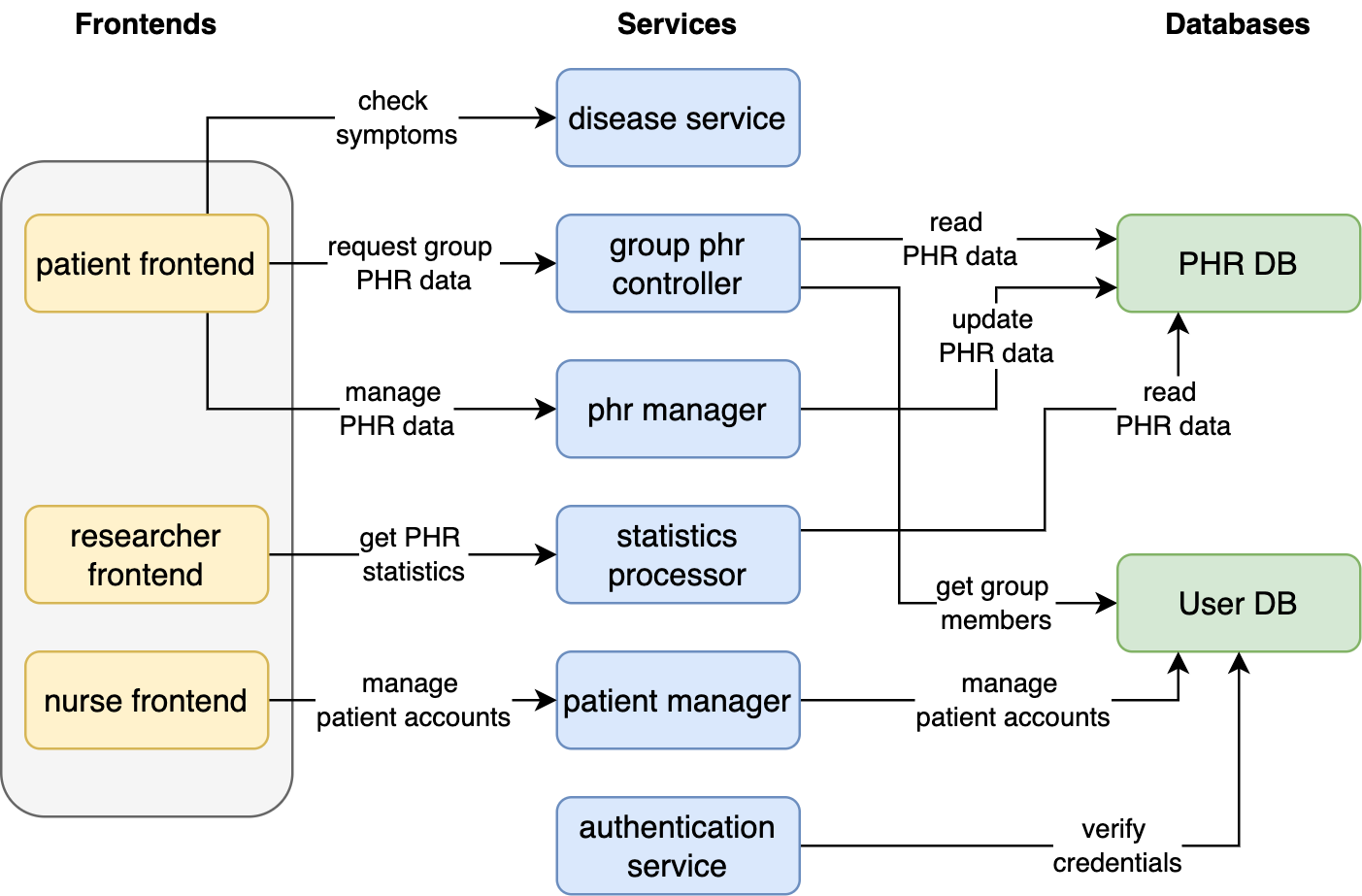}
        \caption{An overview of the patient community example application (adapted from~\cite{Wuyts}). Connections to the authentication service are made from most services, but are left out in the figure for better readability. The frontends are yellow and are all included in one service (indicated by the grey box); the backend services are blue, and databases are green.}
        \label{fig:pce}
    \end{center}
\end{figure}

The architecture of this application has been developed by Wuyts~\cite{Wuyts} as a case example for conducting a LINDDUN analysis.
The Patient Community application is implemented as an open-source project on GitHub\footnote{\url{https://github.com/clouditor/patient-community-example}}. An overview of the components is given in \autoref{fig:pce}.% \url{https://github.com/clouditor/patient-community-example}
The \textbf{frontend} service covers the three frontends described in~\cite{Wuyts}, the patient frontend, the researcher frontend, and the nurse frontend, in one service. It provides the user interface (UI) for these different user groups and is written in TypeScript. Since it is the only component that is used by patients, it is also the entry point for personal data that may be sent to the backend.
% \paragraph{\textbf{auth service}}
The \textbf{auth} service issues authentication tokens for the different users which also encode their roles (nurse, researcher, or patient). It is written in Go.
% \paragraph{\textbf{disease service}}
The \textbf{disease} service allows patients to submit a number of symptoms they experience and returns a list of diseases that typically cause the symptoms. It is written in JavaScript.
% \paragraph{\textbf{phr manager}}
The \textbf{phr manager} allows patients to upload their patient health records (PHR) so they can track their course of disease including what medication they took and which symptoms they experienced. It is written in Python.
% \paragraph{\textbf{group phr controller}}
The \textbf{group phr controller} allows patients to query PHR of their group members, i.e. patients who have the same disease, to compare their course of disease with their own, and compare medications and symptoms. It is written in Python.
% \paragraph{\textbf{nurse api}}
Nurses access the application via the \textbf{nurse api} which is a Java application and allows the registration of new users and their assignment to a group.
% \paragraph{\textbf{statistics service}}
The \textbf{statistics} service is queried by researchers to obtain statistics about existing PHR. It should protect the privacy of patients by aggregating and possibly anonymizing their PHR. To that end, it includes an adapted library that applies $k$-anonymity to the requested data before it is sent to researchers. It is written in Python.
% \paragraph{User DB}
The \textbf{User database} holds patient names and the patients' group assignments. It is a relational PostgreSQL database.
% \paragraph{PHR DB}
The \textbf{PHR database} holds patient health records and is a non-relational MongoDB database.

\subsection{Implemented Weaknesses}
% Also, we can describe what the outcome is a tool should generate when analyzing the application: a certain data flow? A code entity?

Note that we have not implemented all weaknesses that are related to the threats described in~\cite{Wuyts}, since some of them cannot be meaningfully reflected in code, e.g. side-channel threats. % and T16 describes a threat concerning non-compliance of employees.     % threats T14 (information disclosure in internal process), T15 (side-channel attack on internal process), T16 (non-compliance of employees), T18 (non-compliance of management), T20 (content inaccuracy). All other threats are either directly or indirectly present in the application.
% \item Note that the Java microservice makes use of the Spring framework which presents a considerable hurdle in the detection of data flows: many operations like database accesses are abstracted in Spring and are therefore not as straightforward to detect as in the other microservices.

Since personal data must be indicated somehow, we have added comments or decorators where personal data is first introduced, e.g. to variables that hold user input. % Also, we use variable names that can indicate the processing of personal data, e.g. \textit{firstname} or \textit{symptom}, and could be detected by language processing tools.
To enable the correct detection of data flows across services, we also provide a deployment configuration file that specifies, e.g., which parts of the code should be used to build a certain container and where it should be deployed.

In the following we describe some of the weaknesses we have implemented, together with their respective LINDDUN GO names. 
%high-level descriptions of the weaknesses that are implemented in the application ordered by the LINDDUN categories. 
The complete list with more details, like code locations, can be found on the open-source project site on GitHub. 
The main purpose of this list is to provide a clearly documented result list for evaluating privacy analysis tools and methods.

\textbf{Linkability of retrieved data:} The group phr controller (see Figure~\ref{fig:pce}) can access both the User DB and the PHR DB and can therefore link medical records to identifiers.

\textbf{Identifying inbound data:} When users are registered by a nurse, their real first and last names are stored in the User DB.

\textbf{Identifying context:} Weaknesses related to contextual data, like IP addresses, are implicit to the HTTP protocol that is used to transmit PHRs and other data.

\textbf{Non-repudiation of sending:} When users submit PHRs, their submission is logged by the PHR manager. The logs may be used to prove a submission of specific medical data later on.

% \textbf{Non-reputable storage:} Users cryptographically sign messages sent to the backend, where they are stored in a database, making them non-reputable.

\textbf{Detectable credentials:} When a user sends login credentials to the auth service, an HTTP 404 may be returned indicating that the user does not exist. This makes (non-)existing users detectable.

\textbf{Detectable communication:} Data transmissions to the backend can potentially be observed by outsiders.

\textbf{Detectable at storage:} When a user submits PHRs, an arbitrary user ID and group ID can be specified. When the user is not in the specified group, an HTTP 404 may leak information about (non-)existing user-group assignments (i.e. the users' diseases).

\textbf{Detectable at retrieval:} When a user requests group PHRs, an arbitrary user ID and group ID can be specified. When the user is not in the specified group, an HTTP 404 may leak information about (non-)existing user-group assignments (i.e. the users' diseases).

\textbf{No erasure or rectification:} Users can submit their PHRs but there is no possibility implemented for users to rectify or delete their personal data later on.

% \textbf{Disproportionate collection:} 

\textbf{Disproportionate storage:} Nurses register users with their real first and last names. These identifiers, however, are never used which indicates that this data is stored without a proper purpose.

\section{Conclusions and Future Work} % Research Opportunities
We have presented a benchmark application for privacy analysis which can serve as a testing basis for future analysis tools, as well as an educational resource.
% TODO revise number of threats
The application implements privacy weaknesses from all LINDDUN (GO) categories and related to 27 out of the 35 concrete LINDDUN threats. There are, however, many ways privacy weaknesses could be implemented. Evidently, our application does not cover all possible ways that privacy weaknesses can be implemented, but further work should explore different ways of implementation. 
% \item While our application covers many \textit{types} of threats, one limitation is that it only covers few threat implementations per type, e.g. one \textit{detectable at storage} threat (see Appendix~\ref{appendix:a}. Privacy threats, however, can be implemented in a large number of ways, so it is not clear how representative the proposed implementations are for real-world applications.
Also, our application does not include any side-channel weaknesses and other weaknesses that cannot meaningfully be implemented.

Our implementation is composed of microservices that are written in Python, Java, Typescript, Go, and JavaScript. We furthermore include a relational database (PostgreSQL) and a non-relational one (MongoDB). Additionally, we provide deployment information, i.e. a CI/CD script, that specifies which parts of the code form which microservice. 
With our work we hope to inspire researchers and developers to create approaches and tools for the automatic detection of privacy threats, e.g. via static analysis, dynamic analysis, pattern recognition, and others. % We also hope to give the impetus to a discussion about the real-world occurrence and materialization of privacy threats: how should personal data be marked in source code? How can the purpose of personal data collection be defined in a standardized format? Is every transmission of personal data a potential threat?

In future work, we want to extend the application with more weaknesses, as well as synthetic data generation to facilitate real-time testing.
Future work should also analyze in more detail which kinds of analysis approaches, e.g. dynamic analysis, blackbox, whitebox, etc., are suitable to detect certain privacy threats.
Future work should also explore ways to encode the purpose of personal data collection: In contrast to security weaknesses, many privacy threats can not be distinguished from valid data transmissions, because they may be justified by a valid purpose. 
% A standardized format for the definition of the purpose of data collection, however, is not available. Instead, we make certain assumptions or focus on cases that generally present disproportionality threats, e.g. when a datum is collected but not processed or stored at all.
% We therefore suggest to develop a standardized/machine-readable collection purpose form

\section*{Acknowledgement}
This work was partly funded by the European Union Horizon 2020 project MEDINA, Grant No. 952633.

\bibliographystyle{ACM-Reference-Format}
\bibliography{main.bib}

\end{document}